\documentclass[a4paper,twoside,12pt]{article}
\input{obsjkt.sty}

\newcommand{\Msun}{\ensuremath{~{\rm M}_\odot}}                   
\newcommand{\Rsun}{\ensuremath{~{\rm R}_\odot}}                   
\newcommand{\rhosun}{\ensuremath{~\rho_\odot}}                    
\newcommand{\Teff}{\ensuremath{T_{\rm eff}}}                      
\newcommand{\FeH}{\ensuremath{\rm [Fe/H]}}                        
\newcommand{\EBV}{\ensuremath{E(B\!-\!V)}}                        
\newcommand{\Grp}{\ensuremath{G_{\rm RP}}}                        
\newcommand{\degr}{\ensuremath{^\circ}}                           
\renewcommand{\kms}{~km~s$^{-1}$}                                 
\renewcommand{\cd}{~d$^{-1}$}                                     
\newcommand{\mc}[1]{\multicolumn{2}{c}{#1}}                       
\newcommand{\kepler}{\textit{Kepler}}                             
\newcommand{\gaia}{\textit{Gaia}}                                 
\newcommand{\targ}{RZ~Cha}
\newcommand{\targfull}{RZ~Chamaeleontis}

\newcommand{\Msunnom}{\hbox{$\mathcal{M}^{\rm N}_\odot$}}
\newcommand{\Rsunnom}{\hbox{$\mathcal{R}^{\rm N}_\odot$}}
\newcommand{\Lsunnom}{\hbox{$\mathcal{L}^{\rm N}_\odot$}}

\usepackage{rotating}

\begin{document} 

\OBSheader{Rediscussion of eclipsing binaries: \targ}{J.\ Southworth}{2025 April}

\OBStitle{Rediscussion of eclipsing binaries. Paper XXIII. \\ The F-type twin system RZ Chamaeleontis}

\OBSauth{John Southworth}

\OBSinstone{Astrophysics Group, Keele University, Staffordshire, ST5 5BG, UK}


\OBSabstract{\targ\ is a detached eclipsing binary containing two slightly evolved F5 stars in a circular orbit of period 2.832~d. We use new light curves from the Transiting Exoplanet Survey Satellite (TESS) and spectroscopic orbits from \gaia\ DR3 to measure the physical properties of the component stars. We obtain masses of $1.488 \pm 0.011$\Msun\ and $1.482 \pm 0.011$\Msun, and radii of $2.150 \pm 0.006$\Rsun\ and $2.271 \pm 0.006$\Rsun. An orbital ephemeris from the TESS data does not match published times of mid-eclipse from the 1970s, suggesting the period is not constant. We measure a distance to the system of $176.7 \pm 3.7$~pc, which agrees with the \gaia\ DR3 value. A comparison with theoretical models finds agreement for metal abundances of $Z=0.014$ and $Z=0.017$ and an age of 2.3~Gyr. No evidence for pulsations was found in the light curves. Future data from TESS and \gaia\ will provide more precise masses and constraints on any changes in orbital period.}


\section*{Introduction}

The current series of papers \cite{Me20obs} is concerned with determining the physical properties of detached eclipsing binary systems (dEBs) to sufficient precision to be useful for testing the predictions of theoretical stellar models. The intended precision is 2\% or better in the masses and radii of the component stars \cite{Andersen91aarv,Torres++10aarv}, although precisions in the region of 0.2\% can be achieved in the best cases \cite{Maxted+20mn}. Our work uses published spectroscopic radial velocity (RV) measurements combined with new photometry from space missions such as \kepler\ \cite{Borucki16rpph} and TESS \cite{Ricker+15jatis}, which have revolutionised our understanding of binary stars \cite{Me21univ}. 

In this work we turn our attention to the system \targfull\ (Table~\ref{tab:info}), a partially-eclipsing dEB containing two almost identical F-stars on a circular orbit of period 2.828~d. Its variability was discovered by Strohmeier, Knigge \& Ott \cite{Strohmeier++64ibvs} under the moniker of BV~473, from Bamburg photographic patrol plates. Popper \cite{Popper66aj} obtained nine photographic spectra and commented that there were lines of two components with approximately equal intensity. Geyer \& Knigge \cite{GeyerKnigge74ibvs} refined the period to 2.832093(61)~d from $UBV$ observations of most of one eclipse.

J{\o}rgensen \& Gyldenkerne \cite{JorgensenGyldenkerne75aa} presented extensive photometry of \targ\ obtained with the Copenhagen 50~cm telescope sited at ESO La Silla, Chile. They used the Str\"omgren photometer to obtain simultaneous observations in the $uvby$ passbands, totalling 775 points in each band. They fitted the light curves using a rectification procedure \cite{RussellMerrill52book,Gyldenkerne++75aa}, finding the ratio of the radii ($k$) to be close to unity but poorly determined due to the eclipses being partial. They thus fixed $k=1$ to present results for the mean component of the system. The Str\"omgren colour indices were found to be practically the same for the two stars, supporting the imposition of $k=1$ on the light curve solution, and to indicate that they have an approximately solar metallicity ($\FeH = -0.02 \pm 0.15$). The authors also found an effective temperature of $\Teff = 6580 \pm 150$~K for the two stars, and that they had evolved beyond the end of the main sequence.

In an accompanying paper, Andersen et al.\ \cite{Andersen++75aa} (hereafter AGI75) presented photographic spectroscopic observations of \targ\ from which the masses and radii of the mean component were deduced to 1--2\% precision. A modest disagreement was found between the two sets of photographic plates obtained, with the F-series (reciprocal dispersion 20~\AA~mm$^{-1}$) yielding slightly smaller and more uncertain velocity amplitudes than the G-series (12.3~\AA~mm$^{-1}$) plates. AGI75 found the two stars to be almost identical, with a magnitude difference between the spectral line strengths of the components of $0.02 \pm 0.02$~mag (mean error from five spectral lines).           


Giuricin et al.\ \cite{Giuricin+80aa} reanalysed the $uvby$ light curves using the {\sc wink} program \cite{Wood73pasp}, finding that they could differentiate between the two stars. Their results point towards one star being slightly hotter (by 50~K) and also slightly smaller (with $k = 1.061 \pm 0.020$ where the errorbar neglects some sources of uncertainty such as limb darkening). This is plausible in a system where both components are evolved far from the zero-age main sequence. Giuricin et al.\ modelled the four light curves separately and obtained very different results for the $y$ band versus the others (for example a ratio of the radii of 1.40 instead of 1.06), but did not even comment on this discrepancy. The small but detectable difference between the stars was restated by Graczyk et al.\ \cite{Graczyk+17apj}, who included \targ\ in a sample of 35 dEBs constructed to calibrate relations between surface brightness and colour.



\begin{table}[t]
\caption{\em Basic information on \targfull. The $BV$ magnitudes are each the mean of 129 individual measurements \cite{Hog+00aa} distributed approximately randomly in orbital phase. The $JHK_s$ magnitudes are from 2MASS \cite{Cutri+03book} and were obtained at an orbital phase of 0.30. \label{tab:info}}
\centering
\begin{tabular}{lll}
{\em Property}                            & {\em Value}                 & {\em Reference}                      \\[3pt]
Right ascension (J2000)                   & 10 42 24.11                 & \citenum{Gaia23aa}                   \\
Declination (J2000)                       & $-$82 02 14.2               & \citenum{Gaia23aa}                   \\
Henry Draper designation                  & HD 93486                    & \citenum{CannonPickering19anhar2}    \\
\textit{Gaia} DR3 designation             & 5198334162577657984         & \citenum{Gaia21aa}                   \\
\textit{Gaia} DR3 parallax                & $5.7404 \pm 0.0186$ mas     & \citenum{Gaia21aa}                   \\          
TESS\ Input Catalog designation           & TIC 394730113               & \citenum{Stassun+19aj}               \\
$B$ magnitude                             & $8.54 \pm 0.02$             & \citenum{Hog+00aa}                   \\          
$V$ magnitude                             & $8.09 \pm 0.01$             & \citenum{Hog+00aa}                   \\          
$J$ magnitude                             & $7.131 \pm 0.030$           & \citenum{Cutri+03book}               \\
$H$ magnitude                             & $6.941 \pm 0.036$           & \citenum{Cutri+03book}               \\
$K_s$ magnitude                           & $6.904 \pm 0.038$           & \citenum{Cutri+03book}               \\
Spectral type                             & F5~IV-V + F5~IV-V           & \citenum{Andersen++75aa}             \\[3pt]
\end{tabular}
\end{table}


\section*{Photometric observations}


\targ\ has been observed in seven sectors by the NASA Transiting Exoplanet Survey Satellite \cite{Ricker+15jatis,Winn24xxx} (TESS), at a variety of sampling rates. The data from sectors 11, 12 and 13 were obtained at a cadence of 1800~s, from sectors 38 and 39 at 600~s cadence, and from sectors 65 and 66 at both 120~s and 200~s cadence. An eighth set of observations is scheduled in the near future: sector 93 will be observed in June 2025. In this work we concentrate on the data obtained at the highest available cadence.

We downloaded data for all sectors from the NASA Mikulski Archive for Space Telescopes (MAST\footnote{\texttt{https://mast.stsci.edu/portal/Mashup/Clients/Mast/Portal.html}}) using the {\sc lightkurve} package \cite{Lightkurve18}. We specified the quality flag ``hard'' to retain only the best data, and used the simple aperture photometry (SAP) light curves from the SPOC data reduction pipeline \cite{Jenkins+16spie}. The datapoints were converted into differential magnitude and the median magnitude was subtracted from each sector to normalise the data.

We show the resulting light curves in Fig.~\ref{fig:time}. The temporal coverage in the final two sectors, on which we concentrate our efforts, is excellent. A total of 19\,515 and 18\,604 datapoints are available in sectors 65 and 66, respectively.

We queried the \gaia\ DR3 database\footnote{\texttt{https://vizier.cds.unistra.fr/viz-bin/VizieR-3?-source=I/355/gaiadr3}} for all sources within 2~arcmin of \targ. All of the 92 sources returned as a response to our query are fainter than \targ\ by at least 5.01~mag in the \gaia\ \Grp\ passband. We therefore expect the amount of light contaminating the light curve to be negligible. As a confirmation of this, the TICv8 catalogue \cite{Stassun+19aj} indicates that less than 1\% of the light in the TESS light curve of \targ\ may be ascribed to contamination from nearby point sources.


\section*{Light curve analysis}

We combined together the 120-s cadence data of \targ\ from TESS sectors 65 and 66 for a detailed analysis of the photometric variations due to binarity. The primary and secondary eclipses are of similar depth (approximately 0.4~mag) but a difference in depth is apparent on visual inspection. We assigned a time of primary minimum close to the midpoint of the light curve as our reference time of primary minimum ($T_0$) and define star~A to be the star eclipsed at this time. Star~A is therefore hotter than its companion, star~B; it is also the smaller of the two. Their masses are not significantly different (see below).

We modelled the light curve using version 43 of the {\sc jktebop}\footnote{\texttt{http://www.astro.keele.ac.uk/jkt/codes/jktebop.html}} code \cite{Me++04mn2,Me13aa}, fitting for the fractional radii of the stars ($r_{\rm A}$ and $r_{\rm B}$), expressed as their sum ($r_{\rm A}+r_{\rm B}$) and ratio ($k = {r_{\rm B}}/{r_{\rm A}}$), the central surface brightness ratio ($J$), third light ($L_3$), orbital inclination ($i$), orbital period ($P$), and the reference time of primary minimum ($T_0$). Limb darkening (LD) was included in the fit using the power-2 law \cite{Hestroffer97aa,Maxted18aa,Me23obs2}, with the same coefficients used for both stars due to their strong similarity. The linear coefficient ($c$) was fitted and the non-linear coefficient ($\alpha$) was fixed to a suitable theoretical value \cite{ClaretSouthworth22aa,ClaretSouthworth23aa}.

\begin{figure}[t] \centering \includegraphics[width=\textwidth]{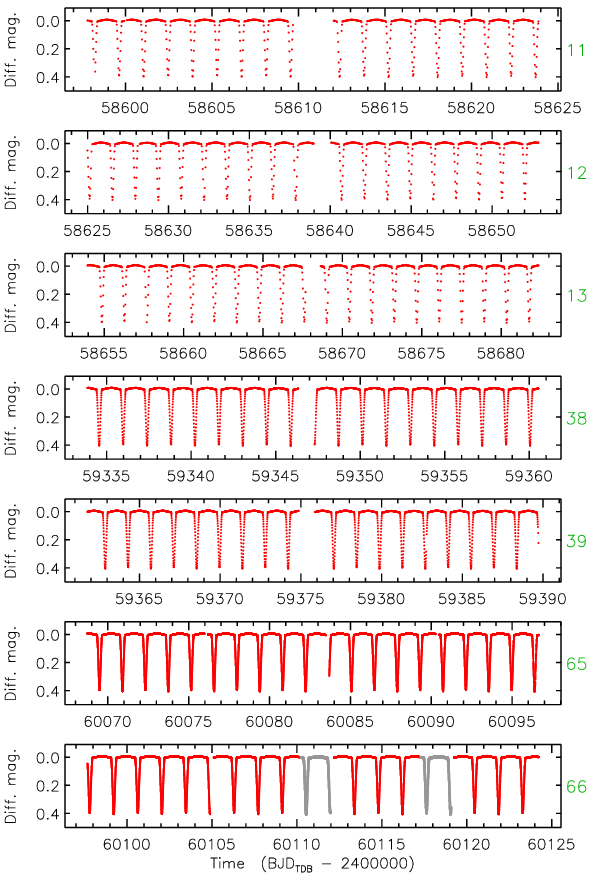} \\
\caption{\label{fig:time} TESS short-cadence SAP photometry of \targ. The flux 
measurements have been converted to magnitude units then rectified to zero magnitude 
by subtraction of the median. Rejected observations are shown as grey open circles. 
The sector number is shown in green to the right of each panel.} \end{figure}

\clearpage

\begin{figure}[t] \centering \includegraphics[width=\textwidth]{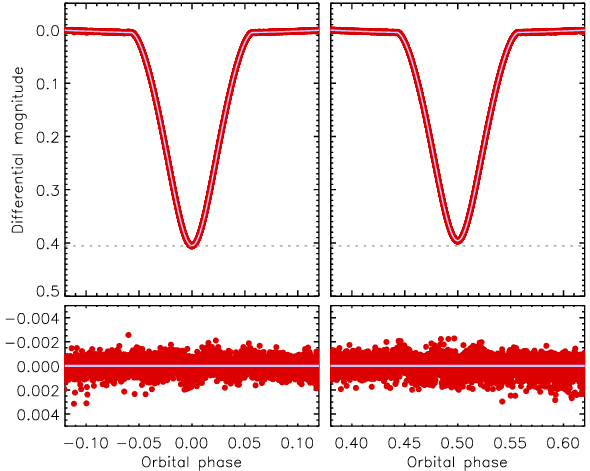} \\
\caption{\label{fig:phase} {\sc jktebop} best fit to the 120-s cadence light curves of \targ\ 
from TESS sectors 65 and 66. The data are shown as filled red circles and the best fit as a 
light blue solid line. A dotted line shows the brightness of the system at the midpoint of 
primary eclipse, and is a visual indicator of the slight difference in depths between the 
two eclipses. The residuals are shown on an enlarged scale in the lower panel.} \end{figure}

After some experimentation it became clear that there were two time intervals where the data had a significantly larger scatter. The affected datapoints were culled from the analysis and are shown in a different colour in Fig.~\ref{fig:time}. It was also apparent that there were slight discontinuities in flux associated with three gaps in the data for each of the two sectors. We therefore applied a total of eight quadratic functions to normalise the out-of-eclipse brightness of the system -- four for each TESS sector. Once these adjustments were made we obtained an excellent fit to the TESS observations (Fig.~\ref{fig:phase}). Our results consistently indicate that star~A is hotter but smaller than star~B.

To obtain errorbars for the fitted parameters we decreased the size of the data errors from the TESS data reduction pipeline to force a reduced $\chi^2$ of unity, then ran the Monte-Carlo and residual-permutation simulations implemented in {\sc jktebop} \cite{Me++04mn2,Me08mn}. The measured parameter values and their errorbars are given in Table~\ref{tab:jktebop}, and in all cases correspond to the residual-permutation values as they are larger than the Monte-Carlo errorbars. 

Our results are in good agreement with previous analyses \cite{JorgensenGyldenkerne75aa,Giuricin+80aa} but with much smaller errorbars. However, our light ratio is slightly inconsistent with the one given by AGI75 from their photographic spectra; the level of disagreement is ambiguous because AGI75 did not specify which star was which in the evaluation of their light ratio.

\begin{table} \centering
\caption{\em \label{tab:jktebop} Photometric parameters of \targ\ measured using 
{\sc jktebop} from the light curves from TESS sectors 65 and 66. The errorbars 
are 1$\sigma$ and were obtained from a residual-permutation analysis.}
\begin{tabular}{lcc}
{\em Parameter}                           &              {\em Value}            \\[3pt]
{\it Fitted parameters:} \\                                                   
Orbital period (d)                        & $       2.8320896  \pm  0.0000013 $ \\
Reference time (BJD$_{\rm TDB}$)          & $ 2460096.389351   \pm  0.000012  $ \\
Orbital inclination (\degr)               & $      83.2920     \pm  0.0060    $ \\
Sum of the fractional radii               & $       0.36509    \pm  0.00012   $ \\
Ratio of the radii                        & $       1.0562     \pm  0.0025    $ \\
Central surface brightness ratio          & $       0.98075    \pm  0.00008   $ \\
Third light                               & $       0.01641    \pm  0.00075   $ \\
LD coefficient $c$                        & $       0.5901     \pm  0.0063    $ \\
LD coefficient $\alpha$                   &            0.4898 (fixed)           \\
{\it Derived parameters:} \\                                                   
Fractional radius of star~A               & $       0.17755    \pm  0.00023   $ \\
Fractional radius of star~B               & $       0.18753    \pm  0.00019   $ \\
Light ratio $\ell_{\rm B}/\ell_{\rm A}$   & $       1.0972     \pm  0.0052    $ \\[3pt]
\end{tabular}
\end{table}



\section*{Orbital ephemeris}

The analysis so far has only used two consecutive sectors of TESS observations, so $P$ and $T_0$ are not as precise as they could be. We therefore fitted each of the TESS sectors individually using {\sc jktebop} to determine times of eclipse. We chose the primary eclipse closest to the midpoint of each sector as best representative of the full sector. We did not obtain any times of secondary eclipse, or times of individual eclipses, as this exceeds the scope of the current work. The times of minimum light used are given in Table~\ref{tab:tmin}

\begin{figure}[t] \centering \includegraphics[width=\textwidth]{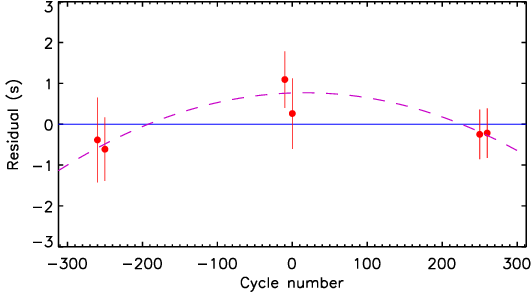} \\
\caption{\label{fig:tmin} Residuals of the times of minimum light from Table~\ref{tab:tmin} 
(red circles) versus the best-fitting ephemerides. The blue solid line and purple dashed 
line indicate residuals of zero for the linear and quadratic ephemeris, respectively. 
Note the extremely small scale on the $y$-axis.} \end{figure}

The orbital ephemeris from the six times of minimum light is
\begin{equation}
  \mbox{Min~I} = {\rm BJD}_{\rm TDB}~ 2459374.206471 (3) + 2.832089764 (13) E
\end{equation}
and the residuals versus the best fit are plotted in Fig.~\ref{fig:tmin}. The times are measured to an extraordinary precision, with a root-mean-square (rms) residual of only 0.56~s. 

We then tried to project the orbital ephemeris back to the times of eclipse given by J{\o}rgensen \& Gyldenkerne \cite{JorgensenGyldenkerne75aa}, including also the time of minimum given by Mallama \cite{Mallama81pasp}. This was unsuccessful because the gap of almost exactly 46~years between our timings and that of Mallama means we cannot confidently assign orbital cycle counts to the older data. We therefore rely on the ephemeris above, which is valid for the duration of the TESS data only.

\targ\ may exhibit low-amplitude period variations. There is a hint of this in our own timings (Table~\ref{tab:tmin}), where the addition of a quadratic term to the ephemeris lowers the rms of the residuals from 0.56~s to 0.27~s (Fig.~\ref{fig:tmin}), and it would also explain our difficulty in adding historical times of minimum to the analysis. Further support for this notion comes from a plot of the residuals versus an orbital ephemeris of \targ\ on the TIDAK website\footnote{\texttt{https://www.as.up.krakow.pl/minicalc/CHARZ.HTM}} \cite{Kreiner++01book}. We leave this matter to the future, where additional insight is expected from the extra sector of data from TESS as well as more extensive compilations of published times of minimum.

\begin{table} \centering
\caption{\em Times of primary eclipse for \targ\ and their residuals versus the fitted ephemeris. \label{tab:tmin}}
\setlength{\tabcolsep}{10pt}
\begin{tabular}{r@{.}lr@{.}lr@{.}lr@{.}l}
\mc{\em Orbital cycle}&\mc{\em Eclipse time (BJD$_{\rm TDB}$)}&\mc{\em Uncertainty (d)}&\mc{\em Residual (d)}\\
   ~~~$-260$ & 0      &      ~~~~~~2458637 & 863128           &   ~~~~0 & 000012       &    ~$-0$ & 000004   \\                                  
      $-250$ & 0      &            2458666 & 184023           &       0 & 000009       &     $-0$ & 000007   \\                                  
      $ -10$ & 0      &            2459345 & 885586           &       0 & 000008       &     $ 0$ & 000013   \\                                  
      $   0$ & 0      &            2459374 & 206474           &       0 & 000010       &     $ 0$ & 000003   \\                                  
      $ 250$ & 0      &            2460082 & 228909           &       0 & 000007       &     $-0$ & 000003   \\                                  
      $ 260$ & 0      &            2460110 & 549807           &       0 & 000007       &     $-0$ & 000003   \\                                  
\end{tabular}
\end{table}

\section*{Radial velocity analysis}

It is important to check the results of the RV analysis presented by AGI75 to ensure consistency with the numbers in the current work. AGI75 noticed an inconsistency between their results from the two series of photographic plates they used, the 20~\AA~mm$^{-1}$ plates giving slightly lower velocity amplitudes ($K_{\rm A}$ and $K_{\rm B}$) than the 12~\AA~mm$^{-1}$ plates. The differences between measurements of the same plates by the various co-authors of the paper using their own methods\footnote{This author confesses he is far too young to have ever used a Grant comparator, although he vaguely remembers seeing one in a store-room at an observatory somewhere.} are smaller both than the inconsistency and the uncertainties. We have collected the various values of $K_{\rm A}$ and $K_{\rm B}$ in Table~\ref{tab:orbit}.

\begin{table} \centering
\caption{\em \label{tab:orbit} Velocity amplitudes measured in different ways for 
\targ. The person who performed the analysis is given in brackets in each case.}
\begin{tabular}{l r@{\,$\pm$\,}l r@{\,$\pm$\,}l}
{\em Source}              & \mc{$K_{\rm A}$ (\kms)} & \mc{$K_{\rm B}$ (\kms)} \\[3pt]
20~\AA~mm$^{-1}$ plates (Imbert)                  & 105.3 & 2.7 & 103.6 & 1.7 \\
20~\AA~mm$^{-1}$ plates (Andersen)                & 106.4 & 0.8 & 106.5 & 1.0 \\
12~\AA~mm$^{-1}$ plates (Gjerl{\o}ff)             & 108.5 & 0.6 & 108.7 & 0.9 \\
12~\AA~mm$^{-1}$ plates (Andersen)                & 108.4 & 0.7 & 107.0 & 0.9 \\
AGI75 adopted value                               & 108.2 & 0.6 & 107.6 & 0.9 \\[3pt]
20~\AA~mm$^{-1}$ plates (this work)               & 106.2 & 1.3 & 105.5 & 1.0 \\
12~\AA~mm$^{-1}$ plates (this work)               & 108.5 & 0.6 & 107.8 & 0.8 \\
20~\AA~mm$^{-1}$ and 12~\AA~mm$^{-1}$ (this work) & 108.0 & 0.6 & 106.7 & 0.7 \\[3pt]
\gaia\ DR3 \texttt{tbosb2}                        & 107.8 & 0.4 & 108.2 & 0.4 \\[3pt]
\end{tabular}
\end{table}

We also extracted the RVs from table~1 of AGI75 to perform our own fits. It is not stated which co-author produced the tabulated RVs, but it is likely to have been Johannes Andersen as he was the only author to analyse both sets of photographic plates. As fitted parameters we specified $K_{\rm A}$, $K_{\rm B}$, the systemic velocity (assumed to be the same for both stars) and a phase offset with respect to our ephemeris in Table~\ref{tab:jktebop}. Uncertainties were calculated using Monte Carlo simulations, and the velocity amplitudes we found are given in Table~\ref{tab:orbit}. 

\begin{figure}[t] \centering \includegraphics[width=\textwidth]{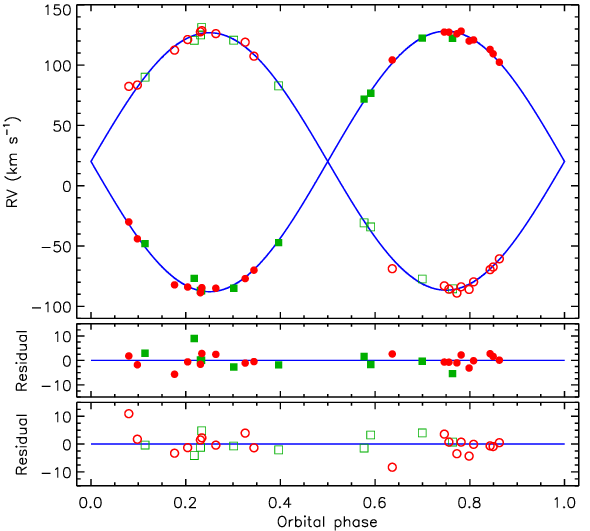} \\
\caption{\label{fig:rv} RVs of \targ\ from AGI75 compared to the best fit from 
{\sc jktebop} (solid blue lines). The RVs for star~A are shown with red filled 
circles for the 20~\AA~mm$^{-1}$ photographic plates and green filled squares 
for the 12~\AA~mm$^{-1}$ plates. The RVs for star~B are shown with red open 
circles for the 20~\AA~mm$^{-1}$ photographic plates and green open squares 
for the 12~\AA~mm$^{-1}$ plates. The residuals are given in the lower panels 
separately for the two components.} \end{figure}

Our first conclusion is that the phase offset is small, hence the primary star adopted by AGI75 is likely the same as our star~A. This conclusion is only valid if changes in the orbital period in the system are small. We fitted the 20~\AA~mm$^{-1}$ RVs, finding an r.m.s.\ scatter of 3.2\kms\ for star~A and 2.5\kms\ for star~B. We then fitted the 12~\AA~mm$^{-1}$ RVs, obtaining scatters of 2.1\kms\ and 3.4\kms\ respectively. With these r.m.s.\ scatters applied as errorbars to the RVs, we then fitted all the AGI75 RVs together (see Fig.~\ref{fig:rv}). The systemic velocities in these fits were all in good agreement, so need not be discussed further. 

\targ\ has also been observed spectroscopically using the RVS instrument \cite{Cropper+18aa} on the \gaia\ mission, and the parameters of its spectroscopic orbit are given in the \texttt{tbosb2} catalogue\footnote{\texttt{https://vizier.cds.unistra.fr/viz-bin/VizieR-3?-source=I/357/tbosb2}} \cite{Gaia23aa2}. Fifteen RVs were automatically measured and fitted for each star; the orbit has the correct period and a very small and negligible \cite{LucySweeney71aj} eccentricity. We verified that the primary star in \texttt{tbosb2} corresponds to our star~A. The velocity amplitudes from \texttt{tbosb2} are given in Table~\ref{tab:orbit}, are consistent with previous determinations, and have a smaller errorbar. It is also clear that $K_{\rm A}$ is smaller and $K_{\rm B}$ is larger than previously found, to the extent that \mbox{$K_{\rm B} > K_{\rm A}$}.

We are thus faced with a choice between adopting the results based on the RVs of AGI75, which are derived from photographic observations and show some inconsistencies, or the orbit given in the \texttt{tbosb2} catalogue based on RVs which are not public and thus cannot be verified. Issues with the \texttt{tbosb2} orbits have previously been noted \cite{Bashi+22mn,Tokovinin23aj,MarcussenAlbrecht23aj,Me23obs4,Me24obs4}, but in the case of \targ\ the orbital parameters are close to the values known from other sources. We have therefore decided to adopt the $K_{\rm A}$ and $K_{\rm B}$ from \texttt{tbosb2}, and note that this can be checked in the near future (late 2026) when \gaia\ DR4 becomes available\footnote{\texttt{https://www.cosmos.esa.int/web/gaia/release}}.

\section*{Physical properties and distance to \targ}

\begin{table} \centering
\caption{\em Physical properties of \targ\ defined using the nominal solar units 
given by IAU 2015 Resolution B3 (ref.~\cite{Prsa+16aj}). \label{tab:absdim}}
\begin{tabular}{lr@{\,$\pm$\,}lr@{\,$\pm$\,}l}
{\em Parameter}        & \multicolumn{2}{c}{\em Star A} & \multicolumn{2}{c}{\em Star B}    \\[3pt]
Mass ratio   $M_{\rm B}/M_{\rm A}$          & \multicolumn{4}{c}{$0.9963 \pm 0.0047$}       \\
Semimajor axis of relative orbit (\Rsunnom) & \multicolumn{4}{c}{$12.109 \pm 0.029$}        \\
Mass (\Msunnom)                             &  1.488  & 0.011       &  1.482  & 0.011       \\
Radius (\Rsunnom)                           &  2.1499 & 0.0058      &  2.2708 & 0.0058      \\
Surface gravity ($\log$[cgs])               &  3.9458 & 0.0018      &  3.8967 & 0.0017      \\
Density ($\!\!$\rhosun)                     &  0.1497 & 0.0007      &  0.1266 & 0.0005      \\
Synchronous rotational velocity ($\!\!$\kms)& 38.41   & 0.10        & 40.57   & 0.10        \\
Effective temperature (K)                   & 6596    & 150         & 6564    & 150         \\
Luminosity $\log(L/\Lsunnom)$               &  0.897  & 0.040       &  0.936  & 0.040       \\
$M_{\rm bol}$ (mag)                         &  2.50   & 0.10        &  2.40   & 0.11        \\
Interstellar reddening \EBV\ (mag)          & \multicolumn{4}{c}{$0.05 \pm 0.02$}			\\
Distance (pc)                               & \multicolumn{4}{c}{$176.7 \pm 3.7$}           \\[3pt]
\end{tabular}
\end{table}


The physical properties of \targ\ were determined using the {\sc jktabsdim} code \cite{Me++05aa} and the results from the analyses described above. The masses are measured to a precision of 0.7\%, are not significantly different from each other, and may be improved once \gaia\ DR4 is published. The radii are measured to a precision of 0.3\%, and star~B is larger and more evolved than star~A. The masses and radii agree well with previous measurements \cite{JorgensenGyldenkerne75aa,Andersen++75aa,Giuricin+80aa}, but are significantly more precise. There is an apparent inconsistency in that the less massive star~B is more evolved than its companion, but the significance of this is too low to be concerning: the mass ratio is only 0.8$\sigma$ below unity.

We adopted a \Teff\ of the system of $6580 \pm 150$~K \cite{JorgensenGyldenkerne75aa} and used the surface brightness ratio and equations from Southworth \cite{Me24obs3} to convert this to individual \Teff\ values (Table~\ref{tab:absdim}). These \Teff\ values were used with the surface brightness calibrations by Kervella et al.\ \cite{Kervella+04aa} and the apparent magnitudes in Table~\ref{tab:info} to determine the distance to the system. A small amount of interstellar reddening of $\EBV = 0.05 \pm 0.02$~mag was needed to align the distances from the optical and infrared passbands. Our best distance estimate is $176.7 \pm 3.7$~pc in the $K_s$-band, which is in agreement with the $174.2 \pm 0.6$~pc from the \gaia\ DR3 parallax.


\section*{Comparison with theoretical models}

\begin{figure}[t] \centering \includegraphics[width=\textwidth]{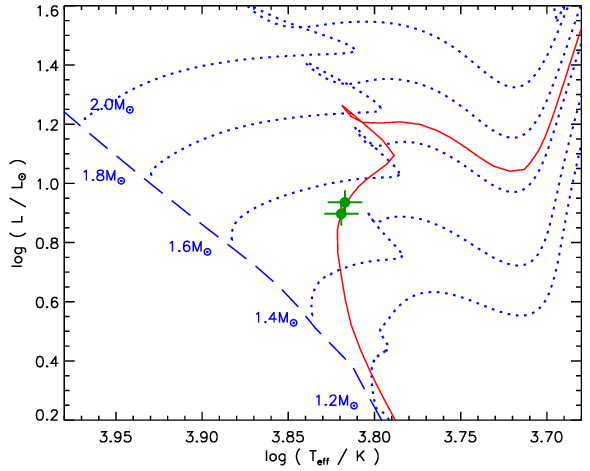} \\
\caption{\label{fig:hrd} Hertzsprung-Russell diagram for the components of \targ\ (filled 
green circles) and the predictions of the {\sc parsec} 1.2S models \cite{Chen+14mn} for 
selected masses (dotted blue lines with masses labelled) and the zero-age main sequence 
(dashed blue line), for a metal abundance of $Z=0.017$. The isochrone for an age of 
2.05~Gyr is shown with a solid red line.} \end{figure}

We compared the properties of \targ\ to the predictions of the {\sc parsec} 1.2S theoretical stellar evolutionary models \cite{Bressan+12mn,Chen+14mn} in the mass--radius and mass--\Teff\ diagrams. We obtained an acceptable fit for a metal abundance of $Z=0.017$ and an age of $2.35 \pm 0.10$~Gyr, in the sense that the theoretical isochrones passed within 1$\sigma$ of the measured properties. A slightly lower metal abundance of $Z=0.014$ gave a better fit for an age of $2.20 \pm 0.10$~Gyr, in that the \Teff\ values were matched almost exactly rather than at the 1$\sigma$ lower errorbar.

J{\o}rgensen \& Gyldenkerne \cite{JorgensenGyldenkerne75aa} found that the components of \targ\ have evolved beyond the main sequence, by comparing their properties with the theoretical models of Hejlesen \cite{Hejlesen80aa,Hejlesen80aas}. AGI75 confirmed the conclusion that both component stars were in the subgiant phase. We investigated this by plotting a Hertzsprung-Russell diagram (Fig.~\ref{fig:hrd}) with the stars and {\sc parsec} evolutionary tracks for a range of masses. This clearly shows that both components are within the main-sequence band, and that an age of 2.05~Gyr is the best match. We find a younger age than in the previous paragraph because we have striven to match \Teff\ and luminosity rather than mass, radius and \Teff. That we find the components to be main-sequence stars rather than subgiants is due to the inclusion of convective core overshooting in more modern theoretical models, which causes the main-sequence band to extend to higher luminosities \cite{MaederMeynet89aa,Stothers91apj}.


\section*{Summary and conclusions}

\targ\ is a dEB containing two F5 stars in a circular orbit of period 2.832~d. We used light curves from the TESS mission and spectroscopic orbits from \gaia\ DR3 to determine the masses and radii of the component stars. With the addition of a published \Teff\ measurement and surface brightness calibrations we determined their luminosities and the distance to the system. The distance we find, $176.7 \pm 3.7$~pc, agrees with the value of $174.2 \pm 0.6$~pc from \gaia\ DR3. The two stars are very similar, having almost identical masses and \Teff\ values, but star~B is larger and thus brighter. We find a mass ratio below unity, in modest disagreement with published values from ground-based spectroscopy, and this result can be checked in the near future when the \gaia\ RVs are published.

Both components are in the upper part of the main-sequence band in the Hertzsprung-Russell diagram, in contrast to previous claims that they have evolved beyond the main-sequence stage. We find acceptable matches to the masses, radii, \Teff\ values and luminosities of the stars for a metal abundance around or slightly below solar, and an age in the region of 2.3~Gyr.

Both components of \targ\ are within the region of the Hertzsprung-Russell diagram where g-mode pulsations can be found \cite{BalonaOzuyar20mn,Mombarg+24aa}, and relatively few g-mode pulsators in dEBs are known \cite{GaulmeGuzik19aa,MeVanreeth22mn}. We therefore checked the TESS light curves for signs of pulsations by fitting them with {\sc jktebop} to remove the signals of binarity, then calculating periodograms using the {\sc period04} code \cite{LenzBreger05coast}. This was done for TESS sectors 11--13, 38 and 39, and 65 and 66. Several possible low-amplitude pulsation frequencies below 3\cd\ were found, but none were consistently present in the three periodograms. A periodogram to the Nyquist frequency of 360\cd\ was calculated for sectors 65 and 66, and showed no significant power beyond 3\cd. We therefore conclude that there is no evidence for pulsations in \targ.


\section*{Acknowledgements}

We are grateful to Dr.\ Pierre Maxted for calculating which star is the primary component in the \gaia\ spectroscopic orbit, and to the anonymous referee for a positive and prompt report.
This paper includes data collected by the TESS\ mission and obtained from the MAST data archive at the Space Telescope Science Institute (STScI). Funding for the TESS\ mission is provided by the NASA's Science Mission Directorate. STScI is operated by the Association of Universities for Research in Astronomy, Inc., under NASA contract NAS 5–26555.
This work has made use of data from the European Space Agency (ESA) mission {\it Gaia}\footnote{\texttt{https://www.cosmos.esa.int/gaia}}, processed by the {\it Gaia} Data Processing and Analysis Consortium (DPAC\footnote{\texttt{https://www.cosmos.esa.int/web/gaia/dpac/consortium}}). Funding for the DPAC has been provided by national institutions, in particular the institutions participating in the {\it Gaia} Multilateral Agreement.
The following resources were used in the course of this work: the NASA Astrophysics Data System; the SIMBAD database operated at CDS, Strasbourg, France; and the ar$\chi$iv scientific paper preprint service operated by Cornell University.



\end{document}